\documentclass{article}%
\usepackage{amsmath}
\usepackage{graphicx}
\usepackage{amsfonts}
\usepackage{amssymb}%
\begin{document}

\title{Problematic solutions to the Landau-Lifshitz equation}
\author{W. E. Baylis and J. Huschilt}
\maketitle

\begin{abstract}
A critical look at the Landau-Lifshitz equation, which has been recently
advocated as an \textquotedblleft exact\textquotedblright\ relativistic
classical equation for the motion of a point charge with radiation reaction,
demonstrates that it generally does not conserve energy-momentum. Its failure
is dramatic in the one-dimensional case of a stepped electric field, where it
predicts discontinuous velocity and thus infinite radiation. The Lorentz-Dirac
equation, on the other hand, in spite of its preacceleration over distances
comparable to the classical electron radius, does not display such problems.

\end{abstract}

\section{Introduction}

Well-known problems of the Lorentz-Dirac (LD)
equation,\cite{Dirac38,LanLif75,Rohrlich90,Parrott87,Yaghjian92,Bay99,Bay76,Bay76a}
such as runaway solutions, preacceleration, nonuniqueness, have led to
proposals of modified classical equations for the relativistic motion of a
point charge in electromagnetic fields, including radiation reaction. Recently
Rohrlich\cite{Rohrlich01} has asserted that forms of the Landau-Lifshitz (LL)
equation\cite{LanLif75,Spohn00} represent the exact equation and should
replace the flawed one (LD)\ derived by Dirac. The purpose of this letter is
to point out that solutions to the LL equation are not generally consistent
with Maxwell's equations for the radiation. The result is a fairly simple to
show but does not seem to have attracted much attention.

The LD equation for the motion of a point charge $e$ of proper velocity $u$ in
an external electromagnetic field $\mathbf{F}$ can be written in the
index-free formulation\cite{Bay99} based on Clifford's geometric algebra of
physical space as%
\begin{equation}
m\dot{u}=e\left\langle \left(  \mathbf{F}+\mathbf{F}_{\mathsf{self}}\right)
u\right\rangle _{\Re}~, \label{LDEq}%
\end{equation}
where the first term on the right $f=e\left\langle \mathbf{F}u\right\rangle
_{\Re}\equiv e\left[  \mathbf{F}u+\left(  \mathbf{F}u\right)  ^{\dag}\right]
/2$ is the covariant Lorentz force, and $\mathbf{F}_{\mathsf{self}}=m\tau
_{e}d\left(  \dot{u}\bar{u}\right)  /d\tau$ is identified as the effective
field of self interaction. We use SI units with $c=1$, dots indicate
derivatives with respect to proper time, and $\tau_{e}$ is two-thirds the time
for light to cross the classical electron radius:%
\begin{equation}
\tau_{e}=\frac{2}{3}\frac{Ke^{2}}{m}\simeq6.266\times10^{-24}~\mathrm{s}%
\end{equation}
with $K=\left(  4\pi\varepsilon_{0}\right)  ^{-1}.$

Equation (\ref{LDEq}) can be expanded in components to give the standard
tensor-component form $m\dot{u}^{\mu}=e\left(  F^{\mu\nu}+F_{\mathsf{self}%
}^{\mu\nu}\right)  u_{\nu},$ where $F_{\mathsf{self}}^{\mu\nu}=m\tau
_{e}d\left(  \dot{u}^{[\mu}u^{\nu]}\right)  /d\tau,$ the summation convention
is adopted, the brackets $[\cdots]$ indicate the antisymmetric part, and the
metric tensor is $\left(  \eta_{\mu\nu}\right)  =\mathrm{diag}\left(
1,-1-1-1\right)  .$ However, the component-free algebraic formulation is
cleaner and offers additional computational tools. In it, paravectors (scalars
plus vectors) represent spacetime vectors. For example $u=\gamma
+\mathbf{u}=u^{\mu}\mathbf{e}_{\mu}$ is the proper velocity with time
component $\gamma\equiv\gamma\mathbf{e}_{0}$ and spatial part $\mathbf{u.}$ An
overbar indicates the Clifford conjugate $\bar{u}=\gamma-\mathbf{u}$ and the
Lorentz-invariant square norm is $u\bar{u}=\gamma^{2}-\mathbf{u}^{2}=u^{\mu
}u^{\nu}\mathbf{e}_{\mu}\mathbf{\bar{e}}_{\nu},$ which gives the Minkowski
spacetime metric $\eta_{\mu\nu}=\left\langle \mathbf{e}_{\mu}\mathbf{\bar{e}%
}_{\nu}\right\rangle _{S}$ as the metric of paravector space, where
$\left\langle x\right\rangle _{S}\equiv\frac{1}{2}\left(  x+\bar{x}\right)  $
is the scalar part of any element $x~.$ Since $u$ is a unit paravector,
$u\bar{u}=1,$ and the proper acceleration $\dot{u}$ is orthogonal to
$u:\left\langle \dot{u}\bar{u}\right\rangle _{S}=0.$ As a consequence,
$\dot{u}\bar{u}$ is a biparavector (a spacetime plane, represented by a
complex vector).

The expansion of $\mathbf{F}_{\mathsf{self}}$ gives%
\begin{equation}
\mathbf{F}_{\mathsf{self}}=m\tau_{e}\frac{d\left(  \dot{u}\bar{u}\right)
}{d\tau}=m\tau_{e}\ddot{u}\bar{u}-P
\end{equation}
where the Lorentz-invariant Larmor power $P=-m\tau_{e}\dot{u}\overline{\dot
{u}}$ is seen from Maxwell's equations $\bar{\partial}\mathbf{F}=\mu_{0}%
\bar{j}$ to be the power radiated by the accelerating point charge. It is
easily seen\cite{LanLif75} that the LD equation (\ref{LDEq}) conserves energy
and momentum with the radiation field between any two points on the world line
of the charge where the acceleration $\dot{u}$ is the same.

The LL equation\cite{LanLif75} is obtained by replacing $m\dot{u}$ in the
radiation term in (\ref{LDEq}) by the Lorentz force $f$%
\begin{align}
m\dot{u}_{\mathrm{LL}}  &  =f+\tau_{e}\left\langle \frac{d}{d\tau}\left(
f\bar{u}\right)  u\right\rangle _{\Re}\nonumber\\
&  =f+\tau_{e}\left(  \dot{f}+\left\langle f\overline{\dot{u}}\right\rangle
_{S}u\right)  \,, \label{LLeq}%
\end{align}
where the second line follows from the reality of $f$ and its orthogonality
with $u.$This is the equation given by Ford and O'Connell\cite{Ford93}, by
Spohn\cite{Spohn00}, and by Rohrlich\cite{Rohrlich01} in his (4a). It is
easily verified that $\dot{u}$ here remains orthogonal to $u.$ As Rohrlich
noted, the last term in (\ref{LLeq}) dictates that the Larmor radiation term
$P$ is replaced in energy-momentum conservation for the LL equation by%
\begin{equation}
P_{\mathrm{LL}}=-\tau_{e}\left\langle f\overline{\dot{u}}\right\rangle _{S}~.
\end{equation}
However, this can differ from the Larmor power $P,$ and it is $P$ that is
given by Maxwell's equations.

The LL equation is the first term in an iterative expansion of the LD equation
in powers of $\tau_{e}$ :%
\begin{equation}
m\dot{u}^{\left(  n+1\right)  }=f+m\tau_{e}\left\langle \frac{d}{d\tau}\left(
\dot{u}^{\left(  n\right)  }\bar{u}\right)  u\right\rangle _{\Re}%
\end{equation}
where $\dot{u}^{\left(  n\right)  }$ is the $n$th-order approximation of
$\dot{u}$ and%
\begin{equation}
\dot{u}^{\left(  0\right)  }=f\,.
\end{equation}
The lowest-order difference between the proper accelerations of the LD and LL
equations is the second-order term $\dot{u}-\dot{u}^{\left(  1\right)
}\approx\tau_{e}^{2}\left(  \dddot{u}+\dot{u}\overline{\dot{u}}\dot{u}\right)
.$

Rohrlich also gives an alternative form [his (4b)], obtained by replacing
$\dot{f}=e\left\langle \mathbf{\dot{F}}u\right\rangle _{\Re}+e\left\langle
\mathbf{F}\dot{u}\right\rangle _{\Re}$ with%
\begin{align}
\dot{f}_{\mathrm{R}}  &  =e\left\langle u\bar{\partial}\right\rangle
_{S}\left\langle \mathbf{F}u\right\rangle _{\Re}+\frac{e}{m}\left\langle
\mathbf{F}\left\langle \mathbf{F}u\right\rangle _{\Re}\right\rangle _{\Re}\\
&  =e\left\langle \mathbf{\dot{F}}u\right\rangle _{\Re}+\frac{1}%
{m}\left\langle \mathbf{F}f\right\rangle _{\Re}.
\end{align}
The difference $\dot{f}-\dot{f}_{\mathrm{R}}$ is first order in $\tau_{e}~.$

\section{A simple example}

As a simple example by which to compare solutions of the equations, consider
one-dimensional motion in a pure electric field $\mathbf{F=E}=E\mathbf{e,}$
where $\mathbf{e}$ is a fixed unit vector. Express the proper velocity in
terms of the rapidity $w$ as $u=\exp\left(  w\mathbf{e}\right)  =\cosh
w+\mathbf{e}\sinh w.$ Then $\dot{u}=\dot{w}u\mathbf{e}$ and the Lorentz force
is $f=e\mathbf{E}u.$The LD equation (\ref{LDEq}) becomes%
\begin{equation}
\dot{w}=\frac{eE}{m}+\tau_{e}\ddot{w}\,, \label{1DLDeqn}%
\end{equation}
whereas the LL equation (\ref{LLeq}) has the form%
\begin{equation}
\dot{w}_{\mathrm{LL}}=\frac{e}{m}\left(  E+\tau_{e}\dot{E}\right)  \,.
\label{1DLLeqn}%
\end{equation}
The LL equation is the first-order iteration of the equation%
\begin{align}
\dot{w}^{\left(  n+1\right)  }  &  =\frac{eE}{m}+\tau_{e}\ddot{w}^{\left(
n\right)  }\,\nonumber\\
\dot{w}^{\left(  0\right)  }  &  =\frac{eE}{m}\\
\dot{w}^{\left(  1\right)  }  &  =\frac{eE}{m}+\tau_{e}\ddot{w}^{\left(
0\right)  }=\frac{e}{m}\left(  E+\tau_{e}\dot{E}\right) \nonumber\\
\dot{w}^{\left(  n+1\right)  }  &  =\frac{e}{m}\sum_{k=0}^{n}\tau_{e}^{k}%
\frac{d^{k}}{d\tau^{k}}E\overset{n\rightarrow\infty}{\longrightarrow}\frac
{e}{m}\left(  1-\tau_{e}\frac{d}{d\tau}\right)  ^{-1}E\nonumber
\end{align}
In the limit $n\rightarrow\infty,$ the iterative solution is seen to approach
the LD solution. To lowest order, the difference between the LD and LL
equations is the second-order term $\dot{w}-\dot{w}^{\left(  1\right)  }%
\simeq\left(  e/m\right)  \tau_{e}^{2}\ddot{E}\,,$ the power difference is
$P_{R}-P\simeq-\left(  e^{2}/m\right)  \tau_{e}^{2}E\dot{E},$ and $\dot
{f}-\dot{f}_{\mathrm{R}}\simeq\left(  e^{2}/m\right)  \tau_{e}uE\dot{E}$, none
of which generally vanishes.

Ford and O'Connell\cite{Ford93} (1993) derive analytical solutions of the LL
equation (\ref{1DLLeqn}) for motion of a charge through an electric field in
the shape of a step:%
\begin{equation}
E\left(  x\right)  =\left\{
\begin{array}
[c]{ll}%
0, & x<0\\
E_{0}, & 0<x<L\\
0, & L<x
\end{array}
\right.  \,. \label{Efield}%
\end{equation}
They also show that these are the smooth limit of numerical solutions for a
smooth rise and fall of the field. Let $\tau=0$ be the proper time that the
charge enters the field from the left ( $x=0$ ) and $\tau=\tau_{1}$ the proper
time that it exits at $x=L.$ Integration of (\ref{1DLLeqn}) gives%
\begin{equation}
w_{\mathrm{LL}}=\left\{
\begin{array}
[c]{cc}%
w_{0}, & \tau<0\\
w_{0}+\varepsilon+\varepsilon\tau/\tau_{e}, & 0<\tau<\tau_{1}\\
w_{2}=w_{0}+\varepsilon\tau_{1}/\tau_{e}, & \tau_{1}<\tau\,,
\end{array}
\right.
\end{equation}
where%
\begin{equation}
\varepsilon=\frac{eE_{0}\tau_{e}}{m}\equiv\alpha\frac{\tau_{e}}{L}\,.
\end{equation}
Note that $w_{\mathrm{LL}}$ is discontinuous: it jumps by $\varepsilon$ as the
charge enters the field at $\tau=0$ and then by $-\varepsilon$ as the charge
leaves at $\tau=\tau_{1}$. Consequently, the acceleration has infinite spikes
as the charge enters and leaves the field. In terms of Dirac delta functions
$\delta\left(  \tau\right)  $ and Heaviside step functions $\theta\left(
\tau\right)  ,$
\begin{equation}
\dot{w}_{\mathrm{LL}}=\varepsilon\left[  \delta\left(  \tau\right)
-\delta\left(  \tau-\tau_{1}\right)  \right]  +\frac{eE_{0}}{m}\theta\left(
\tau\right)  \theta\left(  \tau_{1}-\tau\right)  \mathbf{.}%
\end{equation}
Although $\dot{w}$ and consequently $\dot{u}$ are infinite at $\tau=0,\tau
_{1},$ they are integrable. However, the Larmor radiation, proportional to
$-\dot{u}\overline{\dot{u}}=\dot{w}^{2},$ is not. Thus, according to Maxwell's
equations for the field of a point charge, infinite energy is radiated from
the discontinuities. The distance $x$ traveled in the region $0<x<L$ is
related to $\tau$ by integration%
\begin{equation}
x\left(  \tau\right)  =\int_{0}^{\tau}\sinh w\left(  \tau^{\prime}\right)
\,d\tau^{\prime}=\frac{L}{\alpha}\left[  \cosh\left(  w_{1}+\varepsilon
\tau/\tau_{e}\right)  -\cosh w_{1}\right]
\end{equation}
with $w_{1}=w_{0}+\varepsilon\,.$ In particular, $x\left(  \tau_{1}\right)
=L.$ The energy gain of the charge is $m\left(  \gamma_{2}-\gamma_{0}\right)
,$ where $\gamma_{j}=\cosh w_{j}$ and $w_{2}=\cosh^{-1}\left(  \gamma
_{1}+\alpha\right)  -\varepsilon.$ To second order in $\varepsilon,$%
\begin{align}
\gamma_{2}-\gamma_{0}  &  =\left(  \gamma_{1}+\alpha\right)  \cosh
\varepsilon-\sqrt{\left(  \gamma_{1}+\alpha\right)  ^{2}-1}\sinh
\varepsilon-\gamma_{0}\\
&  \simeq\alpha+\varepsilon\left[  u_{0}-u_{2}^{\left(  0\right)  }\right]
+\varepsilon^{2}\left[  \gamma_{0}+\frac{\alpha}{2}-\frac{\left(  \gamma
_{0}+\alpha\right)  u_{0}}{u_{2}^{\left(  0\right)  }}\right]  ~.
\end{align}
with $u_{2}^{\left(  0\right)  }=\sqrt{\left(  \gamma_{0}+\alpha\right)
^{2}-1}.$

Let's compare this to solutions of the LD equation (\ref{1DLDeqn}). In
numerical solutions, runaways are avoided by integrating backward in
time\cite{Husch76}. In analytical solutions, they are avoided by assuming
$\lim_{\tau\rightarrow\infty}\tau E\left(  \tau\right)  =0$ and putting
$\dot{w}\left(  \infty\right)  =0.$ This gives the usual integral form that
includes brief periods of preacceleration:%
\begin{align}
\dot{w}\left(  \tau\right)   &  =\frac{e}{m}\int_{0}^{\infty}ds\,E\left(
\tau+\tau_{e}s\right)  e^{-s}\\
&  =\left\{
\begin{array}
[c]{ll}%
\left(  e/m\right)  E_{0}e^{\tau/\tau_{e}}\left(  1-e^{-\tau_{1}/\tau_{e}%
}\right)  , & \tau<0\\
\left(  e/m\right)  E_{0}\left(  1-e^{\left(  \tau-\tau_{1}\right)  /\tau_{e}%
}\right)  , & 0<\tau<\tau_{1}\\
0, & \tau_{1}<\tau
\end{array}
\right.  .
\end{align}
A further integration gives%
\begin{equation}
w\left(  \tau\right)  =\left\{
\begin{array}
[c]{ll}%
w_{0}+\varepsilon e^{\tau/\tau_{e}}\left(  1-e^{-\tau_{1}/\tau_{e}}\right)
, & \tau<0\\
w_{0}+\varepsilon\left[  1+\tau/\tau_{e}-e^{\left(  \tau-\tau_{1}\right)
/\tau_{e}}\right]  , & 0<\tau<\tau_{1}\\
w_{2}=w_{0}+\varepsilon\tau_{1}/\tau_{e}, & \tau_{1}<\tau
\end{array}
\right.  .
\end{equation}

\begin{figure}
[tbh]
\begin{center}
\includegraphics[
height=2.0003in,
width=3in
]%
{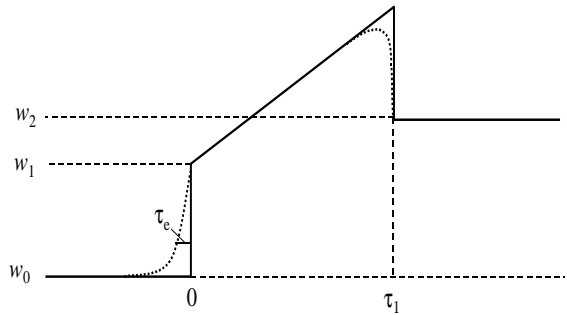}%
\caption{Rapidity $w$ in the case of a stepped field. Solid line: the LL
solution; dotted line: the LD solution. Note that $w$ is continuous in the LD
solution by virtue of the preacceleration over times of about $\tau_{e}$, but
in the LL case, where there is no preacceleration, $w$ is discontinuous. (The
size of $\tau_{e}$ is greatly magnified for clarity.)}%
\end{center}
\end{figure}

\section{Discussion}

Derivations of the LD equation (\ref{LDEq}) generally assume an expansion $u$
in powers of the time difference corresponding to the effective size of the
charge. The limit of vanishing size is then taken, traditionally with mass
renormalization, although such renormalization can be avoided by taking
specified combinations of the self
field\cite{Dirac38,Bay99,Wheeler45,Unruh76,Penrose84}. One cannot expect the
proper velocity to be an analytic function of position in regions where the
field itself is discontinuous. However, discontinuous fields are simply
idealizations convenient for finding analytic solutions. Solutions of both the
LD and LL\ equations can be found numerically for more realistic field
configurations, and they approach the analytic solutions in the appropriate limit.

Rohrlich\cite{Rohrlich01} claims to have derived the LL equation as an exact
classical equation for the point charge. However, his derivation, like most
others\cite{LanLif75}, makes the substitution of an approximate expression
from the Lorentz-force equation with the justification that the radiation term
is small. He then claims that because higher-order derivates of the velocity
than second disappear in Dirac's derivation when the limit of vanishing charge
radius is taken, one should also be able to ignore corresponding derivatives
in the field. This approach appears to argue more forcefully for the
correctness of the LD equation, which as seen above conflicts with the LL
equation. Rohrlich\cite{Rohrlich01} also claims that the LL equation has been
obtained in a rigorous mathematical argument by Spohn\cite{Spohn00}, but Spohn
obtains his critical surface perturbatively and does not claim it to be exact
to all orders of $\tau_{e}$.

The LL equation differs from the LD equation only in second order in $\tau
_{e}$ and its solutions to realistic problems are practically
indistinguishable from those of the LD equation since $\tau_{e}$ is orders of
magnitude smaller than the smallest measurable time interval. Nevertheless, as
seen above, it is inconsistent with Maxwell's equations for the radiation of a
point charge and this inconsistency is dramatic in the case of rectiliner
motion through a stepped field. This is in contrast to the LD equation, which
is consistent.

Yaghjian\cite{Yaghjian92} has proposed a different \textquotedblleft
correction\textquotedblright\ to the LD equation. (Most of his book discusses
a model of the electron as a spherical insulator of finite radius with a fixed
surface charge, but the last section discusses the limit of vanishing radius
to find the motion of a point charge.) He argues that the radiation terms do
not act until the field is turned on and consequently should be multiplied by
a scalar function that approaches a step function in the limit of a point
charge. This eliminates preacceleration. Although his formulation does not
explicitly treat other abrupt changes in the field, for consistency we assume
that the sudden drop in the stepped field has no effect on the radiation terms
until $\tau_{1}$ when the charge leaves the field. However, this prescription
when applied to the stepped field gives precisely the motion of the charge
without \emph{any} radiation reaction. It is therefore also inconsistent with
energy-momentum conservation and Maxwell's equation.

As frequently pointed out\cite{Rohrlich97}, the problems of the LD equation
occur at distance scales well below the Compton wavelength, where quantum
effects become important. Its breakdown in the description of real particles
at such scales is therefore not surprising. Attempts appear so far
unsuccessful to find an alternative classical equation of motion for the point
charge that is free from problems and consistent with energy-momentum
conservation and Maxwell's equations.


\begin{thebibliography}{99}                                                                                               %


\bibitem {Dirac38}P. A. M. Dirac, Proc. R. Soc. A \textbf{167}, 148 (1938).

\bibitem {LanLif75}L. D. Landau and E. M. Lifshitz, \emph{The Classical Theory
of Fields} (4th revised English Edition, translated from the 6th rev. ed. of
the Russian), (Pergamon, New York, 1975).

\bibitem {Rohrlich90}F. Rohrlich, \emph{Classical Charged Particles} (2nd edn,
Addison-Wesley, Reading, MA 1990).

\bibitem {Parrott87}S. Parrott, \emph{Relativistic Electrodynamics and
Differential Geometry} (Springer-Verlag, New York 1987).

\bibitem {Yaghjian92}A. D. Yaghjian, \emph{Relativistic Dynamics of a Charged
Sphere} (Springer, New York, 1992).

\bibitem {Bay99}W. E. Baylis, \emph{Electrodynamics: A Modern Geometric
Approach} (Birkh\"{a}user, Boston 1999), c.12.

\bibitem {Bay76}W. E. Baylis and J. Huschilt, Phys. Rev. D\textbf{13}, 3237 (1976).

\bibitem {Bay76a}W. E. Baylis and J. Huschilt, Phys. Rev. D\textbf{13}, 3262 (1976).

\bibitem {Rohrlich01}F. Rohrlich, Phys. Lett. A \textbf{283}, 276--278 (2001).

\bibitem {Spohn00}H. Spohn, Europhys. Lett. \textbf{50}, 287 (2000).

\bibitem {Ford93}G. W. Ford and R. F. O'Connell, Phys. Lett. A \textbf{174},
182--184 (1993).

\bibitem {Husch76}J. Huschilt and W. E. Baylis, Phys. Rev. D \textbf{13}, 3256 (1976).

\bibitem {Wheeler45}J. A. Wheeler and R. P. Feynman, Rev. Mod. Phys.
\textbf{17}, 157 (1945).

\bibitem {Unruh76}W. G. Unruh, Proc. R. Soc. Lond. A\textbf{348}, 447 (1976).

\bibitem {Penrose84}R. Penrose and W. Rindler, \emph{Spinors and Spacetime,
Vol. 1} (Cambridge University, Cambridge, UK 1984), p. 403.

\bibitem {Rohrlich97}F. Rohrlich, Am. J. Phys. \textbf{65}, 1051 (1997).
\end{thebibliography}
\end{document}